\documentclass[aps,prl,twocolumn,amsmath,amssymb,showpacs]{revtex4-2}

\usepackage[utf8]{inputenc}
\usepackage{graphicx}
\usepackage{amsmath}
\usepackage{amssymb}
\usepackage{siunitx}
\usepackage{xr-hyper}
\usepackage{hyperref}
\usepackage{xcolor}

\newcommand{\ket}[1]{|#1\rangle}

\newcommand{\Er}[1]{\ensuremath{^{#1}\textrm{Er}}}

\begin{document}

\title{Cavity-enhanced optical readout and control of nuclear spin qubits}
\author{Alexander Ulanowski}
\author{Johannes Fr\"uh}
\author{Fabian Salamon}
\author{Adrian Holz\"apfel}
\author{Andreas Reiserer}
\email{andreas.reiserer@tum.de}

\affiliation{Max-Planck-Institut f\"ur Quantenoptik, Quantum Networks Group, Hans-Kopfermann-Stra{\ss}e 1, D-85748 Garching, Germany}
\affiliation{Technical University of Munich, TUM School of Natural Sciences, Physics Department and Munich Center for Quantum Science and Technology (MCQST), James-Franck-Stra{\ss}e 1, D-85748 Garching, Germany}
\affiliation{TUM Center for Quantum Engineering (ZQE), Am Coulombwall 3A, 85748 Garching, Germany}

\begin{abstract}
Their exceptional coherence makes nuclear spins in solids a prime candidate for quantum memories in quantum networks and repeaters. Still, the direct all-optical initialization, coherent control, and readout of individual nuclear spin qubits have been an outstanding challenge. Here, this is achieved by embedding \Er{167} dopants in yttrium orthosilicate in a cryogenic Fabry-Perot cavity, whose linewidth of \SI{65}{\mega\hertz} is much smaller than the \SI{0.9}{\giga\hertz} separation of neighboring hyperfine levels. Frequency-selective emission enhancement thus enables a single-shot readout fidelity of $91(2)\,\%$. Furthermore, a large magnetic field freezes paramagnetic impurities, leading to coherence times exceeding \SI{0.2}{\second}. The combination of nuclear-spin qubits with frequency-multiplexed addressing and lifetime-limited photon emission in the minimal-loss telecommunications C-band establishes \Er{167} as a leading platform for long-range, fiber-based quantum networks.
\end{abstract}   

\maketitle

\section{Introduction}
Quantum networks will enable their users to perform tasks and interact in ways that are impossible using classical devices~\cite{wehner_quantum_2018}, enabling numerous applications in quantum information processing and distributed quantum sensing. First prototypes~\cite{reiserer_colloquium_2022} have demonstrated the entanglement of stationary spin qubits over up to tens of kilometers~\cite{van_leent_entangling_2022, knaut_entanglement_2024, stolk_metropolitan-scale_2024}. Increasing this to even larger distances, however, is hindered by photon loss in optical fibers~\cite{muralidharan_optimal_2016}. It has been proposed that this obstacle can be overcome using quantum repeaters, which require highly efficient interfaces between stationary qubits and optical photons in the minimal loss-band of optical fibers~\cite{holewa_solid-state_2025}, as well as memories with storage times approaching the scale of seconds~\cite{loock_extending_2020}.

Owing to their low sensitivity to electric and magnetic field noise, nuclear-spin qubits are ideally suited to reach this timescale. Previous experiments used atoms trapped in vacuum~\cite{korber_decoherence-protected_2018, wang_single_2021} or embedded as dopants in crystals with a low density of magnetic moments, such as diamond~\cite{dutt_quantum_2007, robledo_high-fidelity_2011, maurer_room-temperature_2012, reiserer_robust_2016, parker_diamond_2024}, silicon~\cite{saeedi_room-temperature_2013}, silicon carbide~\cite{bourassa_entanglement_2020}, and silicate crystals~\cite{zhong_optically_2015, uysal_coherent_2023}. In the latter, the current record coherence---more than ten hours---has been achieved using ensembles of rare-earth dopants~\cite{ wang_nuclear_2025}.

Still, combining such exceptional storage times with efficient and coherent single-photon generation, as required for quantum repeaters~\cite{loock_extending_2020}, is an outstanding challenge. Earlier attempts used nuclear spins around color centers in diamond~\cite{dutt_quantum_2007, neumann_quantum_2010,robledo_high-fidelity_2011} and silicon carbide \cite{lai_single-shot_2024, hesselmeier_high-fidelity_2024}, which were read out indirectly via a controlled coupling to their electronic spin. This way, up to 27 individual nuclear spins have been resolved \cite{abobeih_atomic-scale_2019}. However, in this approach, random flips of the electron reduce the nuclear spin coherence~\cite{maurer_room-temperature_2012, reiserer_robust_2016}, and the indirect nuclear-spin readout via the electronic spin ancilla requires additional electron-nuclear spin quantum gates~\cite{dutt_quantum_2007, neumann_quantum_2010, robledo_high-fidelity_2011, lai_single-shot_2024, hesselmeier_high-fidelity_2024, abobeih_atomic-scale_2019} that are typically implemented by microwave pulses. This reduces the achievable rates and induces heating and imperfections that deteriorate the fidelity. These limitations have been overcome with ensembles of nuclear spins by applying a large magnetic field. When the induced Zeeman splitting strongly exceeds the thermal energy, the electronic spins are frozen to the ground state and decay to it on very fast timescales via the direct spin-lattice relaxation process~\cite{wolfowicz_quantum_2021}. Still, it has been shown that hyperfine interaction can enable all-optical control of nuclear spin ensembles~\cite{rancic_coherence_2018}. Here, we extend this approach to the initialization, coherent control, and optical single-shot readout of individual nuclear-spin qubits.

\begin{figure*}[ht!]
    \centering
    \includegraphics[scale=1.15]{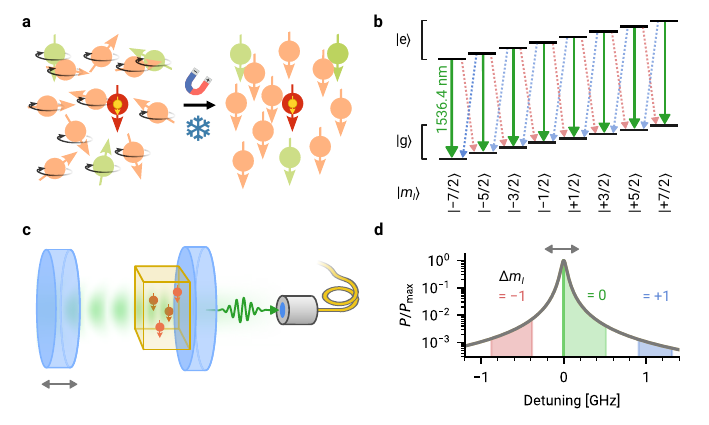}
    \caption{
    \label{fig:Setup_Levels}
    \textbf{Concept of the experiment. a,} Left panel: Quantum bits are encoded in the nuclear spin (gold) of $\Er{167}$ dopants in a crystal. At cryogenic temperatures of $<\SI{2}{\kelvin}$, the electronic spin of the emitters (red) can be aligned, e.g., by optical pumping, eliminating its precession. However, the electronic spins of other dopants and paramagnetic impurities (orange, green) ---potentially unknown and/or without optical transitions--- would be random. Their precession (black arrows) and flipping lead to fluctuating magnetic fields and thus to decoherence. Right panel: When a field of several Tesla is applied, all spins align, which eliminates magnetic field fluctuations even at substantial impurity concentrations. \textbf{b,} \Er{167} dopants exhibit an optical transition at a wavelength of \SI{1536.4}{\nano\meter} (green arrow) between their optical ground $\ket{g}$ and excited $\ket{e}$ state manifolds. Each of them comprises eight nuclear spin levels $\ket{m_I}$, whose energy is shifted by the nuclear Zeeman and hyperfine interactions, such that neighboring levels are separated by up to $\SI{0.9}{\giga\hertz}$ at the applied magnetic field. Using an optical resonator, the nuclear-spin-preserving optical transitions (green arrows) can be enhanced selectively, without altering the decays that increase (red dotted) or reduce (blue dotted) $m_I$ by one quantum. \textbf{c,} Selective enhancement of photon (green curly arrow) emission on the $\Delta{m_I}=0$ transitions requires a resonator with a small mode volume and a very narrow linewidth, $\ll \SI{0.9}{\giga\hertz}$. This is achieved by integrating a $\SI{10.9}{\micro\meter}$ thin membrane of the YSO host crystal (yellow) containing the erbium dopants (red spin symbols) into the optical mode (green) of a frequency-tunable (grey arrow) Fabry-Perot resonator (blue cylinders). \textbf{d,} Via the Purcell effect, the decay rate into the cavity mode on the transition $\ket{-7/2}_e \rightarrow \ket{-7/2}_g$ (green line) is enhanced up to a factor of $P=95(10)$, while detuned optical transitions can exhibit several orders of magnitude lower relative Purcell factors $P/P_\textrm{max}$. The colored areas indicate the spectral bands containing the spin-preserving (light green) and spin-flip (blue and red) transitions.
    }
\end{figure*}

Specifically, we study single erbium dopants in yttrium orthosilicate (YSO) crystals. This combination of emitter and host is particularly promising for quantum networking: First, erbium dopants can emit photons in the telecommunications C-band~\cite{holewa_solid-state_2025}, where loss in optical fibers is minimal. Second, exceptional optical coherence times have been observed in Er:YSO, up to several ms~\cite{bottger_effects_2009}, achieving the lifetime-limit in suited resonators~\cite{merkel_coherent_2020}. Finally, frequency-selective addressing allows coherent control of many individual spin qubits in the same optical mode~\cite{chen_parallel_2020, ulanowski_spectral_2022}---up to several hundreds~\cite{ulanowski_spectral_2024}---paving the way to multiplexed entanglement generation~\cite{ruskuc_multiplexed_2025} over long distances.

While previous experiments in Er:YSO implemented qubits in the electronic spins of the even isotopes~\cite{chen_parallel_2020, ulanowski_spectral_2024}, here we study the nuclear spin of \Er{167}. Similar to the pioneering work with dopant ensembles, in which a second-long coherence was achieved~\cite{rancic_coherence_2018}, we operate at temperatures $<\SI{1.8}{\kelvin}$ in a closed-cycle $^4\mathrm{He}$ cryostat and apply a large magnetic field of $\SI{6.8}{\tesla}$ to fully polarize the electronic spins of both the emitters and of all paramagnetic impurities. As shown in Fig.~\ref{fig:Setup_Levels}a, this reduces intrinsic spin noise and---even in materials with significant impurity concentrations---allows for nuclear spin coherence exceeding seconds~\cite{rancic_coherence_2018} and optical coherence exceeding milliseconds~\cite{bottger_effects_2009}. At the same time, it eliminates optical dephasing caused by superhyperfine interactions~\cite{car_superhyperfine_2020, ulanowski_cavity-enhanced_2025}.

\section{Concept of the experiment}

\begin{figure*}[ht]
    \centering
    \includegraphics[scale=1.15]{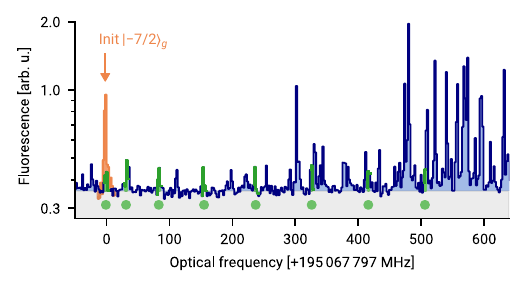}
    \caption{
    \label{fig:Spectrum}
    \textbf{Spectroscopy of single erbium dopants.}
    In pulsed resonant fluorescence spectroscopy, many peaks are observed that originate from individual erbium dopants (blue bars), both with and without hyperfine structure. The shaded gray area below \SI{0.36}{{arb. u.}} indicates the detector dark counts. When applying an optical pumping sequence to initialize an $\Er{167}$ dopant into the $\ket{-7/2}_g$ state, it exhibits a stronger fluorescence intensity on the respective spin-preserving transition $\ket{-7/2}_e \rightarrow \ket{-7/2}_g$ (orange bar). Initialization in the other spin states (not shown) allows assigning all eight spin-preserving transitions of this dopant, highlighted in green in the fluorescence trace and marked by green dots below. A steady increase in frequency is observed from $\ket{-7/2}_e \rightarrow \ket{-7/2}_g$ to $\ket{+7/2}_e \rightarrow \ket{+7/2}_g$ (left to right). 
    }
\end{figure*}

We study site 1 of Er:YSO, which at zero magnetic field exhibits an optical transition frequency of \SI{1536.4}{\nano\meter} between $\ket{g} \equiv Z_1$---the lowest crystal-field level of the $^4I_{15/2}$ ground state---and $\ket{e} \equiv Y_1$---the lowest level of the $^4I_{13/2}$ optically excited state~\cite{bottger_effects_2009}. When applying \SI{6.8}{\tesla} along the b-axis of the crystal, the nuclear spin levels are separated by up to \SI{0.9}{\giga\hertz}~\cite{horvath_extending_2019}, as shown in Fig.~\ref{fig:Setup_Levels}b. This large separation exceeds the inhomogeneous broadening in Er:YSO, which ---even in ensembles~\cite{rancic_coherence_2018}--- enables frequency-selective spin initialization by optical pumping, i.e., repeated excitation on the spin-flip transitions with $\Delta m_I=\pm 1$. Here, $m_I$ denotes the nuclear spin quantum number projection along the magnetic field direction, which ranges from $-7/2$ to $+7/2$ for the $I=7/2$ nuclear spin of $\Er{167}$. Note that the effective $S=1/2$ electronic spin may decay optically to $m_S=1/2$, but in this case returns immediately to the thermal equilibrium state $-1/2$ because of direct spin-lattice relaxation at a rate $\propto B^5$~\cite{wolfowicz_quantum_2021}.

For optical readout, the nuclear spin quantum number $m_I$ needs to be preserved. Albeit this can be ensured upon frequency-selective optical excitation, $m_I$ can change with a significant probability when decaying optically on the $Y_1 \rightarrow Z_1$ transition, as indicated by the blue and red dotted arrows in Fig.~\ref{fig:Setup_Levels}b. Similar to recent experiments with color centers in diamond~\cite{parker_diamond_2024}, the low cyclicity of the transition hinders an all-optical readout with high fidelity. We overcome this challenge by embedding the emitters into an optical Fabry-Perot resonator with a high quality factor of $Q=3\times10^6$, as shown in Fig.~\ref{fig:Setup_Levels}c. The resulting linewidth of \SI{65}{\mega\hertz} (full-width-half-maximum) is much narrower than the separation of the optical transitions between neighboring nuclear spin levels. Thus, as shown in Fig.~\ref{fig:Setup_Levels}d, the decay rate of the detuned spin-flip transitions exhibits no enhancement as $P \ll 1$; it is thus unchanged from its bulk value. In contrast, the cavity selectively enhances the resonant nuclear-spin-preserving transitions ($\Delta m_I = 0$), with a measured Purcell factor~\cite{reiserer_colloquium_2022} of up to 95(10). The resulting increase in cyclicity paves the way for all-optical nuclear-spin qubit readout demonstrated in this work.

\section{Results}

\subsection{Detection of single \Er{167} dopants}

Our experiments use a \SI{10.9}{\micro\meter} thin membrane of YSO that is integrated into a frequency-tunable Fabry-Perot resonator, as described in our earlier works \cite{ulanowski_spectral_2024, ulanowski_cavity-enhanced_2025}. The crystal contains Er in natural isotopic abundance, such that $\sim 23 \%$ of the emitters are of the isotope \Er{167}. At temperatures below \SI{2}{\kelvin}, individual erbium dopants in this crystal exhibit narrow optical transitions down to $\SI{0.2}{\mega\hertz}$~\cite{ulanowski_spectral_2022}, whose center frequencies differ because of local modifications in the crystal environment, e.g., strain, defects, or other dopants. These random frequency shifts allow frequency-multiplexed addressing of individual emitters~\cite{chen_parallel_2020, ulanowski_spectral_2022}. Co-doping with Eu allows for a tailored decrease of the spectral density~\cite{ulanowski_spectral_2024}.

To identify a suited \Er{167} emitter, we employ pulsed fluorescence spectroscopy with laser pulses of \SI{8}{\micro\second} duration that are frequency-chirped over \SI{2}{\mega\hertz}. Using superconducting nanowire single-photon detectors, we detect the emitted fluorescence photons after the laser is turned off. We then scan the excitation laser frequency while keeping the cavity on resonance. While the even isotopes without nuclear spin lead to a strong fluorescence at a single resonance frequency, one expects eight smaller peaks---one for each nuclear spin ground state---for the spin-preserving transitions of \Er{167}. An example spectrum showing this signature can be seen in Fig.~\ref{fig:Spectrum}. The spin-flip transitions, which are detuned by a few hundred MHz, do not give a measurable signal (not shown) because their Purcell enhancement is reduced in proportion to the branching ratio; in addition, in the case of repeated probing at the same frequency, the signal will decay because of unintended optical spin pumping.

\subsection{Purcell-enhanced optical spin initialization}

In the following, we will focus on the optical transition between $\ket{-7/2}_g$ and $\ket{-7/2}_e$. First, the transitions between the outermost spin states ($m_I=\pm7/2$) are expected to exhibit the largest cyclicity, which is favorable for spin readout. Second, the spin can be initialized in $\ket{-7/2}_g$ by driving all the $\Delta m_I=-1$ optical transitions, as demonstrated previously with atomic ensembles \cite{rancic_coherence_2018}. However, the use of a single emitter in an optical resonator calls for an adaptation of the protocol, which we will describe in the following. 

To initialize the spin in $\ket{-7/2}$ starting from a random state, one needs to de-populate all other levels---a procedure termed optical pumping. To this end, one can selectively excite all seven transitions of the red sideband, i.e., all spin-flip transitions that decrease the quantum number of the nuclear spin by one, $\Delta m_I=-1$. The subsequent decay from the excited state can randomly change the spin. Under continuous or repeated driving, the system will undergo many optical excitations and decays until it ends up in $\ket{-7/2}_g$, from which no excitation on the red sideband is possible.

The timescale of the optical pumping process is governed by the emitter lifetime and by the branching ratio of the spin-flip versus the spin-preserving decays. Thus, the presence of the resonator enables a speed-up~\cite{reiserer_colloquium_2022}, as the Purcell effect enhances the branching of the spin-preserving lines and leads to a faster optical decay, reducing the excited-state lifetime of \SI{11.4}{\milli\second} \cite{bottger_spectroscopy_2006} down to \SI{0.12}{\milli\second}.

The spectral dependence of the relative Purcell factors $P/P_\textrm{max}$ of the individual transitions are shown in Fig.~\ref{fig:Setup_Levels}d for a situation in which the resonator is tuned precisely to $\ket{-7/2}_e \rightarrow \ket{-7/2}_g$, such that the decay of this transition is enhanced. The Purcell factors are smaller for the higher spin states, and negligible for the $\ket{+7/2}_e \rightarrow \ket{+7/2}_g$ transition. To achieve the fastest initialization, all seven red-sideband transitions could be excited simultaneously. Instead, for technical reasons, we choose to apply them sequentially, such that the required pulses can be generated from a single laser that is shifted in frequency by an electro-optical modulator.

In the experiment, we use pulses of $\SI{20}{\micro\second}$ duration that are frequency-chirped over \SI{10}{\mega\hertz}. We start at the $\ket{+7/2}_g \rightarrow \ket{+5/2}_e$ frequency, then proceed with $\ket{+5/2}_g \rightarrow \ket{+3/2}_e$ and continue analogously until $\ket{-5/2}_g \rightarrow \ket{-7/2}_e$. Repeating this sequence leads to a strong increase of the population in $\ket{-7/2}_g$, as can be seen in Fig.~\ref{fig:Spectrum}, where the fluorescence signal on the corresponding spin-preserving transition is strongly increased (orange). The initialization fidelity can be characterized by autocorrelation measurements. To achieve the highest value, up to \SI{97.3(9)}{\percent}, we repeat the pumping sequence 500 times in the experiments on single-shot readout that will be described below.

While owing to the branching ratios, the initialization in the outermost spin states, such as $\ket{-7/2}_g$, is expected to be most efficient, the spin can also be initialized in other states by replacing red-sideband transitions with the corresponding ones on the blue sideband. This allows us to identify the precise frequencies of all $\Delta m_I=0,\pm1$ optical transitions of the studied dopant, 22 in total, and thus to fully characterize the energy levels of the ground and excited state. The assigned spin-preserving transitions in the spectrum are marked with green dots at the bottom of the panel in Fig.~\ref{fig:Spectrum}. 

\subsection{Single-shot nuclear spin qubit readout}
After implementing a pulse sequence that allows for efficient initialization, we now turn to the optical readout of the nuclear spin state. To this end, we define the qubit in the two lowest levels in the ground state, $\ket{-7/2}_g$ and $\ket{-5/2}_g$, and apply optical excitation pulses on resonance with the spin-preserving transition $\ket{-7/2}_g \rightarrow \ket{-7/2}_e$. As this transition is far detuned from all other optical transitions, this allows for a spin-selective excitation, such that the detection of a photon after a narrowband excitation pulse unambiguously heralds the $\ket{-7/2}_g$ state. However, absence of a photon does not herald the $\ket{-5/2}_g$ state because of the finite excitation and detection probabilities. Thus, several optical excitations are needed to achieve an unambiguous readout of the nuclear spin. This is only possible if the transition exhibits a high cyclicity, which means that the spin state is not changed in the optical decay. In our experiment, this is enabled by the selective resonator enhancement. Furthermore, the resonator improves the photon detection efficiency to $11(1)\,\%$ and reduces the optical lifetime, which leads to a speed-up of the readout process. Only this combination of advantages makes a single-shot readout feasible.

\begin{figure}[ht]
    \centering
    \includegraphics{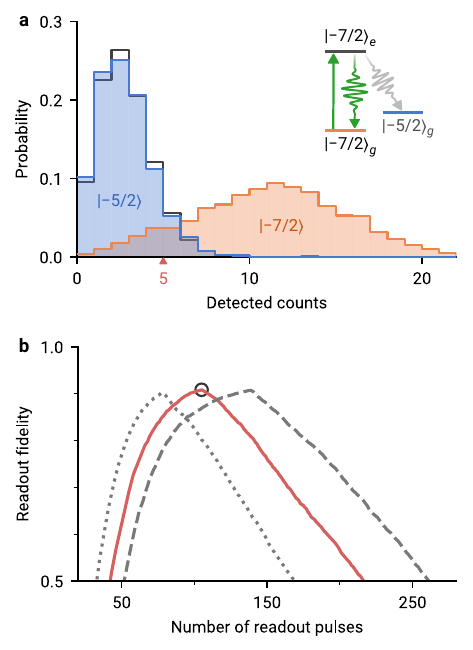}
    \caption{\label{fig:Init_Readout}
    \textbf{Optical nuclear-spin qubit readout. a,}
    Inset: The qubit is encoded in the two lowest-energy states $\ket{-7/2}_g$ (orange) and $\ket{-5/2}_g$ (blue) of the ground state $\ket{g}$, and the resonator is tuned to selectively enhance the optical readout transition $\ket{-7/2}_g \leftrightarrow \ket{-7/2}_e$ (green). After initialization, a readout is performed by measuring the fluorescence photons (green curly arrow) after resonant laser excitation (straight arrow). Main panel: After 110 readout pulses, the number of detector photons differs depending on the qubit initialization. The distributions are clearly separated. When assigning the $\ket{-7/2}_g$ nuclear spin state in case at least $n=5$ photons are detected, an average readout fidelity of $91(2)\,\%$ is achieved. This value is limited by photons detected when the spin is prepared in $\ket{-5/2}_g$, with a distribution (blue) that is dominated by detector dark counts that are observed even without excitation laser pulses (black open bars). \textbf{b,} The achieved fidelity depends on the number of readout pulses and the threshold photon number $n$ (dotted line: 4, solid red line: 5, dashed line: 6). The optimal value (black circle) is found at 110 pulses and $n=5$ (as shown in panel a).}
\end{figure}

For a first experimental demonstration, after spin initialization, we repeatedly excite the spin-preserving $\ket{-7/2}_g\rightarrow{\ket{-7/2}_e}$ transition and measure the fluorescence. We use pulses of \SI{8}{\micro\second} duration that are frequency-chirped over \SI{2}{\mega\hertz}. If the spin is in the $\ket{-7/2}_g$ state, several photons can be detected in each readout attempt. If, however, the spin is in $\ket{-5/2}_g$, or in another state outside of the qubit manifold, one expects to detect only a small number of fluorescence photons that can originate from off-resonant driving or from detector dark counts. The histograms obtained after 110 excitation pulses are shown in Fig.~\ref{fig:Init_Readout}a. Clearly, the distribution obtained when initializing the spin in $\ket{-7/2}_g$ (orange) is different from that of $\ket{-5/2}_g$ (blue), which both are close to binomial distributions with average values of 10.69(5) and 2.356(24) photons, respectively. Setting the discrimination threshold at the optimal value of $n=5$ detected photons gives a single-shot readout fidelity of $91(2)\,\%$. Here, we define fidelity as the minimum probability of correctly assigning the nuclear spin states $\ket{-7/2}_g$ and $\ket{-5/2}_g$, and we do not correct for initialization errors.

The histograms in Fig.~\ref{fig:Init_Readout}a were measured at the optimal pulse number and discrimination threshold. This is the result of a systematic optimization, shown in Fig.~\ref{fig:Init_Readout}b. From the histograms, it can be seen that the main factor that limits the achieved fidelity is detector dark counts at a rate of $\SI{43.9(1)}{\hertz}$, as can be seen from the distribution of detection events in the absence of optical excitation pulses (black open bars), which is very close to the distribution in the dark state $\ket{-5/2}_g$ (blue bars). As the dark counts can be reduced by more than an order of magnitude at comparable efficiency in state-of-the-art devices \cite{mueller_free-space_2021}, we expect that it will be possible to increase the single-shot-readout fidelity above $98\,\%$, surpassing common thresholds for topological quantum error correction~\cite{fowler_surface_2012}.

\subsection{All-optical coherent control}

After demonstrating the single-shot readout of individual nuclear spins, we now turn to their coherence properties. We start by measuring the lifetime, which is $\SI{33(3)}{\second}$ at the used temperature of $\SI{1.72(1)}{\kelvin}$ and magnetic field of $\SI{6.8}{\tesla}$. Further improvement may be achieved at lower temperatures.

\begin{figure*}[ht!]
    \centering
    \includegraphics[scale=1.15]{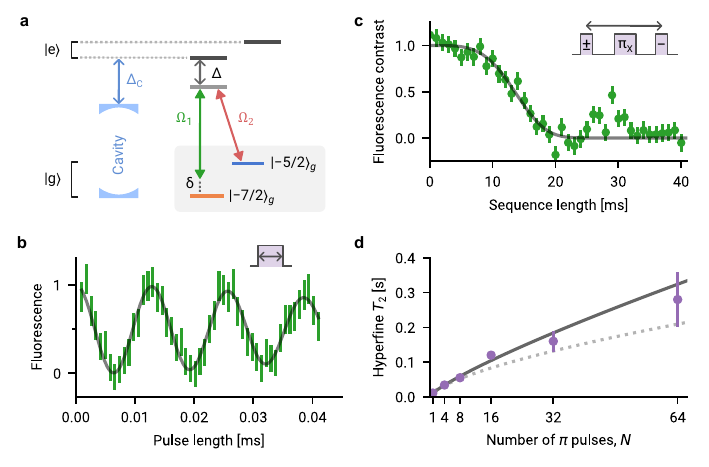}
    \caption{\label{fig:Coherent_Control}
    \textbf{All-optical coherent control and hyperfine coherence. a,}
    For coherent control of the nuclear-spin qubit (gray box), the resonator is temporarily detuned by $\Delta_\textrm{C}=\SI{-400}{\mega\hertz}$ from the $\ket{-7/2}_g\leftrightarrow \ket{-7/2}_e$ frequency. Then, Raman transitions are driven using square laser pulses with Rabi frequencies $\Omega_1$ and $\Omega_2$ that are irradiated simultaneously with a frequency difference that matches the qubit transition when $\delta=0$. Scattering is avoided by choosing a detuning of $\Delta = -\SI{90}{\mega\hertz}$ from the readout transition. 
    \textbf{b,} When varying the Raman pulse duration, Rabi oscillations are observed at $\delta = 0$. A cosine function with a stretched-exponential envelope (solid line) fits the data well. \textbf{c,} In a Hahn-Echo experiment, consisting of a pulse sequence of $\pi/2 - \pi - \pi/2$ (see inset), a coherence time of $T_\mathrm{Hahn}=\SI{14.8(9)}{\milli\second}$ is obtained from a fit to a stretched-exponential function (solid line). \textbf{d,} The coherence time can be further extended by dynamical decoupling, in which $N$ equidistant $\pi$-pulses are applied sequentially. To reduce the influence of pulse errors, the phase of the pulses is altered between rotations around the $X$ and $Y$ axes, forming an XY$(N)$ sequence. With this, the coherence time increases with the number of pulses (purple data) according to $T_\mathrm{DD} \propto N^{0.82(2)}$ (black fit curve) up to a value of $T_\mathrm{DD}=\SI{0.28(8)}{\s}$. The dotted gray line shows the expected increase for a slowly varying spin bath, $\propto N^{2/3}$.}
\end{figure*}

To determine the coherence time, we implement coherent control over the nuclear spin states. Because of their small gyromagnetic ratio, direct driving with radio-frequency fields would require powers of many Watts to achieve Rabi frequencies in the kHz regime, leading to heating and resonator instability. Therefore, we instead implement all-optical control. Specifically, we drive Raman transitions with two control fields, as sketched in Fig.~\Ref{fig:Coherent_Control}a. To avoid that scattering from the excited state limits the spin control fidelity, while still operating at moderate drive powers of $\SI{10}{\milli\watt}$, we chose a detuning of $\Delta = -\SI{90}{\mega\hertz}$ from the $\ket{-7/2}_g\leftrightarrow \ket{-7/2}_e$ transition. 

To ensure that both fields have a comparable and constant amplitude, and to make the Rabi frequency insensitive to residual frequency fluctuations of the resonator, we tune the latter such that both fields exhibit equal detunings of opposite sign from the cavity resonance. Then, we irradiate pulses of varying duration to observe Rabi oscillations between the states of the qubit manifold, as shown in Fig.~\ref{fig:Coherent_Control}b. The oscillations are damped because of power and/or polarization fluctuations of the Raman laser pulses. Dephasing can be excluded for pulse durations below $\SI{0.1}{\milli\second}$, as the dephasing time is measured as $T_2^*=\SI{0.62(3)}{\milli\second}$ in a Ramsey sequence. This value exceeds the $\SI{56(9)}{\nano\second}$ coherence of the electronic spins of erbium in the same host~\cite{cova_farina_coherent_2021} by more than four orders of magnitude, demonstrating the low sensitivity of the nuclear spin qubits to magnetic field fluctuations.

\subsection{Nuclear-spin qubit coherence}

To further increase the coherence, we use a Hahn-echo sequence. As shown in Fig.~\ref{fig:Coherent_Control}c, we find a Hahn-echo time of $T_{\text{Hahn}}=\SI{16.7(9)}{\milli\second}$. This value is much smaller than that found in ensemble measurements on site 2 in \Er{167}, exceeding one second~\cite{rancic_coherence_2018}. This reduction, and the revival of $T_\text{Hahn}$ around \SI{30}{\milli\second}, originate from fluctuations of the bias magnetic field that can be reduced in the future by active damping. 

Alternatively, the coherence can be extended by dynamical decoupling. To this end, we apply $\pi$-pulses, alternating around the $X$ and $Y$ axes to reduce the sensitivity to pulse errors~\cite{suter_colloquium_2016}. With this, we find a coherence time up to $T_\mathrm{DD}=\SI{0.28(8)}{\second}$ after 64 pulses. Eliminating residual pulse imperfections will likely enable extending this even further. The scaling of the coherence time with the number of applied pulses $N$ follows a power law, $T_\mathrm{DD} \propto N^{0.82(2)}$, as shown in Fig.~\ref{fig:Coherent_Control}d, in good agreement with the expectation from the fluctuating magnetic field in the setup, and different from the power law $\propto N^{2/3}$ expected for a slowly varying spin bath. 

\section{Discussion}

The above measurements demonstrate the all-optical initialization, coherent control, and single-shot readout of nuclear spin qubits in a situation where all electronic spins in the sample are frozen to the ground state. Our technique eliminates three main sources of decoherence that have plagued earlier experiments on erbium and other Kramers rare-earth dopants: First, their large sensitivity to fluctuating magnetic fields caused by paramagnetic impurities in the host material. Second, their anisotropic interaction with other dopants of the same type, which cannot be decoupled effectively~\cite{merkel_dynamical_2021}. Finally, the superhyperfine interaction of erbium dopants with the Y nuclei of the host, which leads to a fast collapse of the optical coherence~\cite{car_superhyperfine_2020} that can be avoided at large magnetic fields~\cite{bottger_effects_2009, ulanowski_cavity-enhanced_2025}.

While the achieved coherence is already sufficient for the implementation of quantum repeater protocols~\cite{loock_extending_2020}, it can be further extended by eliminating the current limitation, classical magnetic-field noise in our cryostat. Alternatively, the fidelity of the Raman pulses may be improved by active pulse-area stabilization, which would facilitate the application of more pulses to increase the coherence. However, a limit may be reached because of the small probability of spontaneous decay after populating the optically excited state during the Raman pulses. This may hinder achieving the lifetime limit of $\SI{66(6)}{\second}$. If such ultralong coherence times are desired, it will be advantageous to switch to a crystal with a low abundance of nuclear spins, such as calcium tungstate, in which the coherence of the electronic spin at mK temperatures is four orders of magnitude longer~\cite{le_dantec_twenty-threemillisecond_2021} than that achieved in YSO~\cite{merkel_dynamical_2021}.

In spite of the optical scattering, our demonstrated use of Raman pulses for optical control of individual spins has the advantage that it enables comparably low drive powers, on the order of \SI{10}{\milli\watt}, most of which ($>90\,\%$) is reflected because of the detuning from the cavity resonance. This results in much lower heating compared to the tens of Watts that are required for dynamical decoupling using radio-frequency fields applied via external coils~\cite{gundogan_solid_2015, ortu_storage_2022}. In addition, the Raman fields induce an AC Stark shift to the dopants that depends on both the detuning and the field amplitude, and thus the emitter location in the resonator mode. Therefore, the technique could enable selective addressing of individual spins, similar to earlier experiments with electronic spins in Er:YSO that used microwave driving and an additional AC Stark shift laser~\cite{chen_parallel_2020}.

While the achieved fidelity of the single-shot readout is still below that required for state-of-the-art quantum error correction protocols~\cite{breuckmann_quantum_2021}, our analysis reveals that fidelities exceeding $98\,\%$ could be directly achieved by photon detectors with lower dark counts, which in turn can be realized when efficiently filtering out remaining blackbody radiation~\cite{shibata_ultimate_2015}. In addition, a higher cyclicity can be obtained by increasing the cooperativity of the system via a reduced resonator mode waist, or---potentially---by using a vector magnet to apply the magnetic field in a direction in which the nuclear-spin-flip probability is reduced. At the same time, tuning the magnetic field to a specific operating point can also make the ground-state or optical transition frequencies insensitive to magnetic field fluctuations~\cite{mcauslan_reducing_2012}, which would further extend their coherence times.

\section{Outlook}

Already without these improvements, our system is well-suited for the optically-mediated generation of entanglement between remote erbium dopants. Compared to recent realizations with ytterbium~\cite{ruskuc_multiplexed_2025}, the increased outcoupling efficiency of our setup (up to $76(5)\,\%$), its operation at telecommunications frequencies, and its combination of lifetime-limited optical~\cite{merkel_coherent_2020} and \SI{0.2}{\second} long ground-state coherence even without $^3\text{He}$ refrigerators make nuclear spins of erbium a prime candidate for quantum repeater experiments~\cite{loock_extending_2020}. This could be based on the presented implementation of a nuclear spin qubit combined with time-bin encoding of photonic qubits~\cite{bernien_heralded_2013, uysal_spin-photon_2025}. Alternatively, one could make use of the larger Hilbert space of the nuclear spin manifold to encode quantum information in small Schrödinger cat states. In contrast to recent implementations of this idea with donor qubits~\cite{yu_schrodinger_2025}, erbium allows for optical interfacing of such states, opening a path towards efficient error-correction protocols based on spin-cat codes~\cite{omanakuttan_fault-tolerant_2024}.

However, our approach is not restricted to \Er{167}:YSO, the studied combination of host and dopant; it may also be implemented in other physical systems, e.g., Er:Si~\cite{yang_zeeman_2022, gritsch_optical_2025, fruh_spectral_2026}, $\text{Er:Y}_2\text{O}_3$, or color centers in diamond, silicon, and silicon carbide. This requires resonators with a linewidth that is smaller than the hyperfine coupling and the spectral diffusion of the emitters. Not only Fabry-Perot cavities, as studied in this work, but also nanophotonic resonators~\cite{asano_photonic_2018, gritsch_optical_2025} may be suited to this end. Thus, we expect that the demonstrated combination of long memory time, efficient photon extraction, photon coherence, spectral multiplexing capacity, and telecom operation opens unique perspectives for large-scale quantum networks in several hardware platforms.

\section{Funding}
Funded by 1) the Deutsche Forschungsgemeinschaft (DFG, German Research Foundation) under the German Universities Excellence Initiative - EXC-2111 - 390814868 and via the individual grant agreement 547245129, 2) by the Munich Quantum Valley, which is supported by the Bavarian state government with funds from the Hightech Agenda Bayern Plus, and 3) by the European Union (ERC project OpENSpinS, number 101170219). Views and opinions expressed are those of the authors only and do not necessarily reflect those of the European Union or the European Research Council. Neither the European Union nor the granting authority can be held responsible for them.

\bibliographystyle{apsrev4-2}
\bibliography{bibliography.bib}

\end{document}